\documentclass[aps,prl,superscriptaddress,twocolumn]{revtex4-2}

\usepackage[utf8]{inputenc}
\usepackage[version=3]{mhchem}
\usepackage{amsmath}
\usepackage{amsfonts}
\usepackage{braket}
\usepackage{url}
\usepackage[colorlinks=true, citecolor=blue,allcolors=blue]{hyperref}
\usepackage[draft]{graphicx}
\usepackage{color}
\usepackage{float}
\usepackage[export]{adjustbox}
\usepackage{wrapfig}
\usepackage{array,multirow}
\usepackage{tabularx}
\usepackage{pdfpages}
\usepackage{pgffor}

\makeatletter
\renewcommand\frontmatter@abstractwidth{\dimexpr\textwidth-1in\relax}
\AtBeginDocument{\let\LS@rot\@undefined}
\makeatother

\begin{document}

\title{Generation of Bright and Controllable Isolated Attosecond X-Ray Pulses from Synchronized Mid-Infrared and Ultra-short Ultraviolet Laser Fields}
\date{\today}

\author{Davis Robinson}
\email{davis.robinson@ucf.edu}
\affiliation{Department of Physics, University of Central Florida, Orlando, FL 32816, USA}

\author{Kyle A. Hamer}
\affiliation{Department of Physics, University of Central Florida, Orlando, FL 32816, USA}

\author{Chelsea Kincaid}
\affiliation{The Ohio State University, Columbus, OH 43210, USA}

\author{Michael Chini}
\affiliation{The Ohio State University, Columbus, OH 43210, USA}

\author{Nicolas Douguet}
\email{nicolas.douguet@ucf.edu}
\affiliation{Department of Physics, University of Central Florida, Orlando, FL 32816, USA}

\begin{abstract}
We investigate, by solving the time-dependent Schr\"{o}dinger equation in the single-active-electron approximation in helium, a two-color scheme for tabletop high-order harmonic generation (HHG) that combines a mid-infrared (MIR) driving field with an ultrashort ultraviolet (UV) pulse that could be generated via resonant dispersive wave emission in gas-filled hollow-core fibers. This configuration enables the generation of bright, isolated, and tunable attosecond X-ray pulses. In contrast to single-color driving schemes, which suffer from low conversion efficiency, unfavorable wavelength scaling, and limited spectral control, the MIR+UV approach provides a practical and controllable route for advancing tabletop ultrafast spectroscopy and real-time molecular imaging within current experimental capabilities.
\end{abstract}

\maketitle


The generation of attosecond pulses has opened the door to probing real-time electron dynamics in matter, unveiling ultrafast phenomena that were previously inaccessible \cite{kling2008, krausz2009, altucci2010, ramasesha2016, mauger2025}. While large-scale free-electron laser (FEL) facilities~\cite{pellegrini2016,ratner2019,huang2021} have driven key advances in ultrafast science~\cite{duris2020,grell2023,wang2025}, their complexity and limited accessibility hinder broader use. In contrast, table-top setups using high-harmonic generation (HHG) \cite{corkum1993, lewenstein1994} offer a practical and widely-accessible platform for attosecond pulse generation \cite{antoine1996, paul2001, hentschel2001}, making them essential to the continued development of ultrafast science. HHG has enabled isolated attosecond pulses through amplitude- and polarization-gating techniques~\cite{gouliemakis2008,sola2006} and, more recently, using the macroscopic effects of an intense 2-$\mu$m driver~\cite{chen2014}. Control over the structure of the harmonic plateau has also been achieved with few-cycle, carrier–envelope-phase (CEP)-stabilized pulses~\cite{ishii2014}. In parallel, extending the driving wavelength into the mid-infrared (MIR) regime has pushed the harmonic spectra into the water window~\cite{li2017, ren2018}, broadening attosecond spectroscopy to biological systems in their native aqueous environments~\cite{trabattoni2019, jordan2020, zinchenko2023}.

Despite their successes, single-color driving schemes remain fundamentally limited. A single laser pulse governs both electron emission and the subsequent dynamics leading to recombination \cite{corkum1993, lewenstein1994}, restricting one's ability to independently control or shape the emission process. Additionally, HHG suffers from inherently low conversion efficiency, especially when using long-wavelength drivers \cite{shiner2009, yavuz2012, austin2012, frolov2015}. Therefore, achieving simultaneous control over the spectral range and pulse duration remains a significant challenge, especially in the X-ray regime \cite{takahashi2004, jin2018, johnson2019}. 

Consequently, various multi-color driving schemes have been proposed. For example, combining a single-color laser pulse with its second~\cite{perry1993, kim2004, mauritsson2006}, third~\cite{watanabe1994, liang2006, kroh2018}, or higher~\cite{ishikawa2003} harmonics, phase-locked to the fundamental driving field, has been shown either to extend and tune the harmonic plateau or to significantly enhance the signal. Other approaches have employed attosecond pulse trains (APTs) to trigger the emission step of HHG \cite{schafer2004, biegert2006, ishikawa2007, gademann2011, heldt2023}, demonstrating increased yield and control via selective excitation of a single quantum path. Very recently, Piper {\it et al.}~\cite{piper2025} reported experimental sub-cycle control of electron emission by combining an extreme ultraviolet (XUV) APT with a delayed NIR field. However, while promising, these approaches remain limited by the inherently-low photon flux of high-harmonic sources and the experimental challenges associated with controlling the seed harmonic pulse train. Crucially, none of them yet constitute a single method capable of producing bright, isolated attosecond pulses in the X-ray regime.

Recent advances in laser technology, particularly the development of novel ultraviolet (UV) sources based on soliton dynamics and resonant dispersive wave (RDW) emission in gas-filled hollow-core fibers \cite{travers2019}, offer a promising route toward high-power optical attosecond pulse generation, with the potential to overcome current limitations in photon flux and temporal–spectral control. These sources provide high-energy, tunable (100–400 nm), few-femtosecond (1–5 fs) UV pulses, offering a powerful route to enhance tabletop attosecond generation and to enable applications in molecular imaging ~\cite{zuo96,xu09,pullen15} and pump–probe spectroscopy. In this work, we harness these advances by combining an intense, ultrashort UV pulse with a MIR driving field to generate high-order harmonics in atomic helium gas. We theoretically demonstrate that this configuration enhances control over the temporal structure, spectral content, and signal strength of the emitted radiation, enabling the generation of bright, isolated, and tunable attosecond pulses in the soft X-ray regime.

\begin{figure}[!t]
  \centering
  \includegraphics[width=\linewidth]{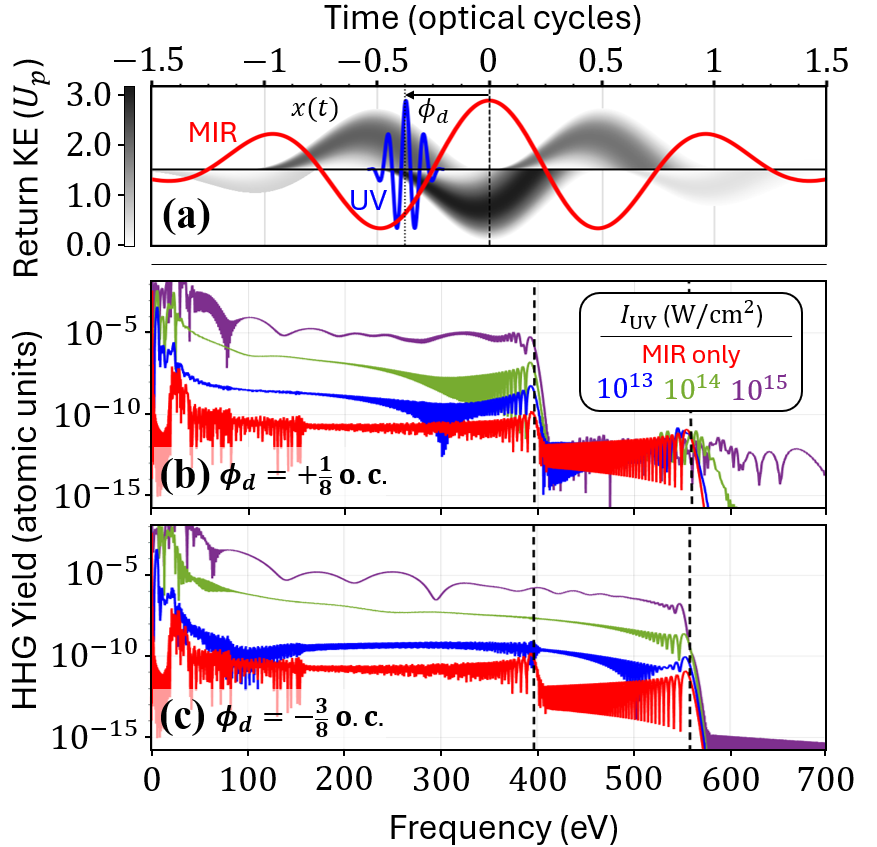}
  \caption{(a) Schematic of our two-color MIR+UV scheme: the delay between the MIR (red) and UV (blue) pulses is denoted by $\phi_{d}$, and the semiclassical electron trajectories $x(t)$ are shown throughout the MIR pulse. (b,c) UV-intensity-dependent HHG spectra for (b) $\phi_{d} = +1/8$ o.c.\ and (c) $\phi_{d} = -3/8$ o.c.. The two plateaus, calculated from the semiclassical three-step model \cite{corkum1993, lewenstein1994} and labeled i and ii, are positioned at $2.22U_{p} + I_{p}$ and $3.17U_{p} + I_{p}$, respectively.}
  \label{fig:1}
\end{figure}

Our proposed two-color scheme is shown in Fig.~\ref{fig:1}(a): a 3-$\mu$m pulse (red) with an intensity of $2 \times 10^{14}$~W/cm$^2$ is combined with a 244-nm UV pulse (blue) with a variable intensity. Both pulses use a sine-square envelope with a total duration of four optical cycles, corresponding to full-width at half-maximum (FWHM) values of approximately 20 fs for the MIR pulse and 1.6 fs for the UV pulse. The frequency of the UV pulse ($\approx$ 5 eV) is such that five UV photons are just enough to ionize helium in the absence of the MIR field. The phase delay between the two pulses, expressed in optical cycles of the MIR, is denoted as $\phi_{d}$. A negative value of $\phi_{d}$ implies that the UV pulse arrives before the MIR pulse. 

In Fig.~{\ref{fig:1}(a)}, we also show the classical electron trajectories $x(t)$ emitted at different times in the driving MIR pulse. According to the three-step model \cite{schafer1993, chang2016}, the electron is first released with zero initial velocity, then accelerated by the MIR field, and finally driven back to recombine with the parent ion, releasing its excess energy as high-harmonic radiation. The color contrast of the trajectories indicates the electron’s kinetic energy at the moment of recombination, as shown by the color bar on the left of Fig.~\ref{fig:1}(a). 

In Figs.~\ref{fig:1}(b) and \ref{fig:1}(c), we show HHG spectra, computed by solving the time-dependent Schr\"{o}dinger equation (TDSE) in the single-active-electron (SAE) approximation, for two different values of $\phi_{d}$ and four different UV intensities. Further details about these simulations can be found in the supplemental material (SM) \cite{supplemental}. The MIR-only curve (red) is identical for both panels. In the absence of the UV pulse, the HHG spectrum exhibits a two-plateau structure, due to the envelope effects that are present for a few-cycle MIR pulse: the first plateau, extending out to $2.22U_{p} + I_{p}$, originates mostly from electrons emitted just after the main peak of the MIR field at $t=0$. Here, $U_p$ denotes the ponderomotive energy and $I_p$ the ionization potential of helium. The second plateau, extending out to $3.16U_{p}+I_{p}$, comes from electrons released in the previous half-cycle (just after $t=-0.5$~o.c.). We refer to the global maximum of the MIR field at $t=0$ as the main peak, and to the preceding maximum as the side peak. The cutoff of each plateau follows the predictions of the classical model. The much higher tunneling probability at the main peak relative to the side peak causes the difference in HHG yield between the two plateaus. Near the cutoff energies of both plateaus, we see oscillations as a function of energy, arising from the interference between short and long trajectories that recombine with the same electron energy but a different phase.

Next, we analyze the HHG process driven by the combined MIR and UV fields, using the time delay $\phi_{d}$ as a control parameter for electron emission. For $\phi_{d} = +1/8$~o.c.\ [Fig.~\ref{fig:1}(b)], the UV pulse predominantly enhances electrons ionized just after the main peak of the MIR field, thereby strengthening only the first plateau. By contrast, for $\phi_{d} = -3/8$~o.c.\ [Fig.~\ref{fig:1}(c)], ionization at the side peak of the MIR field is strongly increased, which enhances the signal across the second plateau and effectively removes the two-plateau structure. As expected, the overall enhancement of the HHG yield is governed by the UV intensity, with the detailed scaling analyzed in more details below. Notably, increasing the UV intensity from $10^{13}$ to $10^{15}$~W/cm$^{2}$ boosts the yield by more than six orders of magnitude.

\begin{figure*}[!t]
  \centering
  \includegraphics[width=\linewidth]{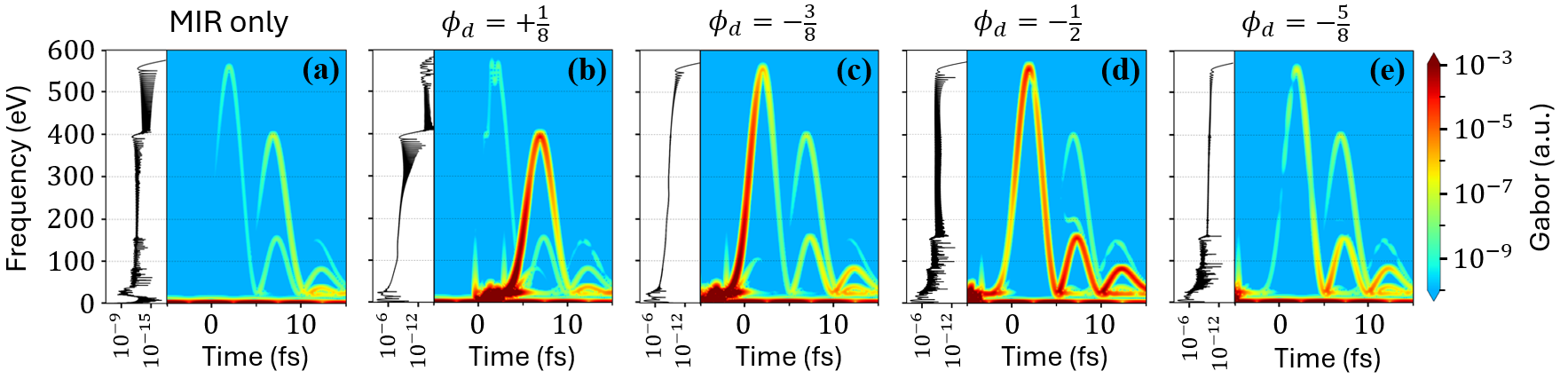}
  \caption{High-harmonic spectra (left) and Gabor transforms (right) of the dipole acceleration for (a) 3$\mu$m MIR-only, and for MIR+UV at four different values of the phase delay $\phi_{d}$: (b) $+1/8$, (c) $-3/8$, (d) $-1/2$, and (e) $-5/8$, all in optical cycles.}
  \label{fig:2}
\end{figure*}

Additional structures can be identified in the HHG spectra of Figs.~\ref{fig:1}(b) and (c). With increasing intensity, the interference fringes fade at $\phi_d=-3/8$~o.c., indicating diminished contrast between the contributions from short and long trajectories. At higher UV intensities, slow energy-dependent oscillations emerge in the HHG spectrum, originating from the sub-cycle structure of the UV pulse near the tunneling regime. Furthermore, for a delay of $\phi_{d} = +1/8$~o.c., the plateau cutoff extends because electrons released at the preceding side peak acquire additional energy through their interaction with the UV field just before recombination.

To analyze the structure of the HHG signal in greater detail, we performed a Gabor transform of the MIR-only signal and of the MIR+UV signals at selected phase delays, as shown in Fig.~\ref{fig:2}. The Gabor transform of the MIR-only case in Fig.~\ref{fig:2}(a) clearly shows the temporal buildup of the two-plateau structure (the corresponding HHG spectrum is presented in the inset) and indicates that the dominant contribution to the harmonic signal originates from long trajectories, i.e., electrons that tunnel near the peak of the field and recombine at least 0.5~o.c.\ later. An additional feature, with a maximum energy of about 150~eV around $7$~fs, results from a second rescattering of long trajectories, where electrons recombine during the subsequent half-cycle, producing a signal of comparable strength to the first recombination.

The Gabor transforms of the MIR+UV signals were computed for several phase delays at a UV intensity of $10^{14}$~W/cm$^{2}$. For all delays, we observe an enhancement of the signal by several orders of magnitude (note the logarithmic color scale in Fig.~\ref{fig:2}). For $\phi_{d} = +1/8$~o.c.\ [Fig.~\ref{fig:2}(b)], only the first plateau is enhanced, predominantly through short trajectories. An extension and distortion of the second plateau cutoff is also visible, originating from the effect of the UV field near the electron recombination time, as discussed previously. In Fig.~\ref{fig:2}(c), the UV pulse is shifted by half a cycle relative to Fig.~\ref{fig:2}(b), which enhances the second plateau exclusively via short trajectories, leading to the complete disappearance of interference in the corresponding HHG spectrum. The enhancement of the signal is highly localized in time and no increase from multiple-recollision trajectories is observed. This behavior stands in sharp contrast to the case of $\phi_{d} = -1/2$~o.c.\ in Fig.~\ref{fig:1}(d), which predominantly selects long trajectories and results in a global enhancement of the multiple-rescattering contributions. The distinction arises because only long trajectories can give rise to such multiple recombinations \cite{lewenstein1994, he2014}. Beyond HHG, this suggests that the MIR+UV scheme can serve as a filter of multi-rescattering contributions, which is highly desirable for molecular imaging, e.g., via light-induced electron diffraction (LIED) \cite{xu2012,biegert2021}, where suppressing background from these trajectories would improve the retrieval of structural and temporal information \cite{degiovannini2023}. Finally, for $\phi_{d} = -5/8$~o.c.\ [Fig.~\ref{fig:2}(e)], most of the electrons ionized by the UV pulse cannot recombine but instead drift away, so that the Gabor transform strongly resembles that of the MIR-only case in Fig.~\ref{fig:2}(a).

\begin{figure}[t]
  \centering
  \includegraphics[width=\linewidth]{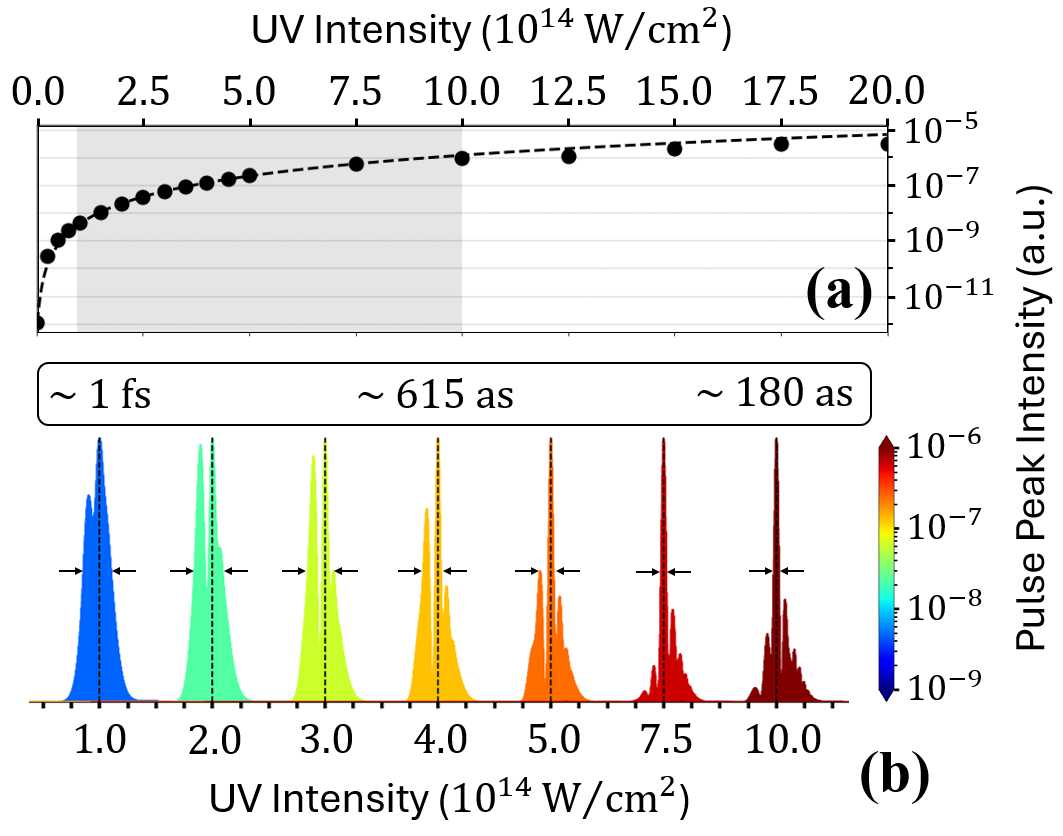}
  \caption{(a) Peak intensity of the attosecond pulse as a function of the UV intensity. The trendline for intensity below $3.5\times10^{14}$W/cm$^2$ is proportional to $I_{\textsc{uv}}^{\alpha}$, with $\alpha = 2.44$. (b) Normalized isolated attosecond pulses generated via UV-enhanced HHG with increasingly-intense few-cycle UV pulses, synchronized with the MIR such that the short trajectories are enhanced. The colormap indicates the peak intensity of the attosecond pulse, and the arrows the decreasing FWHM. }
  \label{fig:4}
\end{figure}

In Fig.~\ref{fig:4}, we examine how the peak intensity and FWHM of the generated attosecond pulse scale with UV intensity in the two-color scheme. To produce bright isolated attosecond pulses in the water window~\cite{li2017, ren2018}, we select $\phi_{d}$ near $-3/8$~o.c., as it enhances the complete plateau, up to $3.16U_{p} + I_{p}$ (see Fig.~\ref{fig:1}(c)). By selecting short trajectories, we also avoid unfavorable phase-matching effects in macroscopic propagation \cite{antoine1996, balcou1997, gaarde2002, mairesse2003}. To extract the attosecond pulse from the generated harmonic spectrum, we filter the dipole acceleration to remove frequencies below 200~eV. As a reference, the main attosecond pulse generated with the MIR field alone has an intensity of $1.1 \times 10^{-12}$ a.u.\ and a FWHM duration of about 1.9~fs. Using the UV pulse described above, even with a modest intensity of $10^{14}$ W/cm$^{2}$, the intensity of the isolated attosecond pulse increases to $4.8 \times 10^{-9}$ a.u.\ (due to the enhancement of the emission yield) and the FWHM duration decreases to approximately 980~as (due to the selectivity of the UV pulse).

In the perturbative multi-photon regime, ionization driven by the UV alone proceeds via a five-photon step, implying a peak intensity scaling as $I_{\text{peak}}\propto I_{\textsc{uv}}^{\alpha}$ with $\alpha=5$. 
By contrast, our calculations below $5\times 10^{14}\,\mathrm{W/cm^{2}}$ in Fig.~\ref{fig:4}(a) yield $\alpha\simeq 2.44$. This reduced exponent arises because, on the UV timescale, the quasi-static MIR field lowers the barrier by 
$\Delta V\simeq 2\sqrt{|E_{\textsc{mir}}(\phi_d)|}$~\cite{freeman1978,ni2008}, 
where $E_{\textsc{mir}}(\phi_d)$ is the MIR electric field at the UV peak. 
At this delay, the effective ionization potential is reduced to $I_p^{\rm eff}\approx 0.42~\mathrm{a.u.}$, 
corresponding to an effective two- to three-photon process once the UV bandwidth and MIR nonadiabaticity are taken into account. 
As shown in the SM~\cite{supplemental}, SFA calculations confirm that the peak-intensity exponent decreases with increasing MIR intensity, consistent with the barrier-suppression picture. 
At intensities above $5\times10^{14}\,\mathrm{W/cm^{2}}$, however, the data deviate from this simple scaling as the UV-driven dynamics transition into the tunneling regime and ultimately to over-the-barrier ionization.

As shown in Fig.~\ref{fig:4}(b), the FWHM of the attosecond pulse shrinks from about $1$~fs below $I_{\textsc{uv}} = 3.5 \times 10^{14}$~W/cm$^{2}$ to $180$~as at $7.5 \times 10^{14}$~W/cm$^{2}$, and remains nearly constant thereafter. In the multi-photon regime, the attosecond pulse duration is governed by the UV envelope. With the Keldysh parameter approaching $\gamma\simeq 1$ near $I_{\textsc{uv}}\simeq 10^{15}$~W/cm$^{2}$, the ionization becomes confined to the field maxima, reducing the temporal gate to the sub-cycle scale set by the UV period. In the tunneling regime, and neglecting the MIR field during the emission step, the ADK model~\cite{ammosov1986} gives (see SM~\cite{supplemental}) an instantaneous tunneling rate near a field maximum,
\begin{equation}
W(t)\simeq W_{0}\exp\left(-t^2/2\sigma^{2}\right),
\end{equation}
with $W_0$ the rate at the peak ($t=0$) and the width
\begin{equation}\label{eq:width}
\sigma\simeq\omega^{-1}_{\textsc{uv}}\sqrt{3E_{\textsc{uv}}/2\kappa^{3}}, \qquad
\kappa=\sqrt{2I_{p}} .
\end{equation}
At $I_{\textsc{uv}}=10^{15}$~W/cm$^{2}$, this yields a FWHM  electron-emission gate of $\Delta t_e\simeq 64$~as, and a corresponding recombination time of $\Delta t_r\simeq 2.5\,\Delta t_e$, such that $\Delta t_r\approx 160$~as, in quantitative agreement with the numerical results. Beyond this intensity, the attosecond pulse grows slowly, $\sigma \propto I_{\textsc{uv}}^{1/4}$. We repeated the calculation for a long, multi-cycle MIR pulse (FWHM = 40 fs) and again found isolated attosecond pulse generation, in contrast to the MIR-only case~\cite{supplemental}.

The above formula provides only a crude estimate of the attosecond pulse duration, as it neglects peak distortions induced by the MIR field, intrinsic attochirp, macroscopic propagation, CEP, and multi-cycle structure effects. In particular, Eq.~(\ref{eq:width}) predicts a decrease of the attosecond pulse duration with increasing UV frequency. Although this trend appears in our simulations, the pulse shape is complicated by ionization from multiple nearby cycles. Further work will be needed to identify optimal conditions for generating shorter attosecond pulses; however, it is promising that such a simple model can be used to predict these effects.

\begin{figure}[!b]
  \centering
  \includegraphics[width=\linewidth]{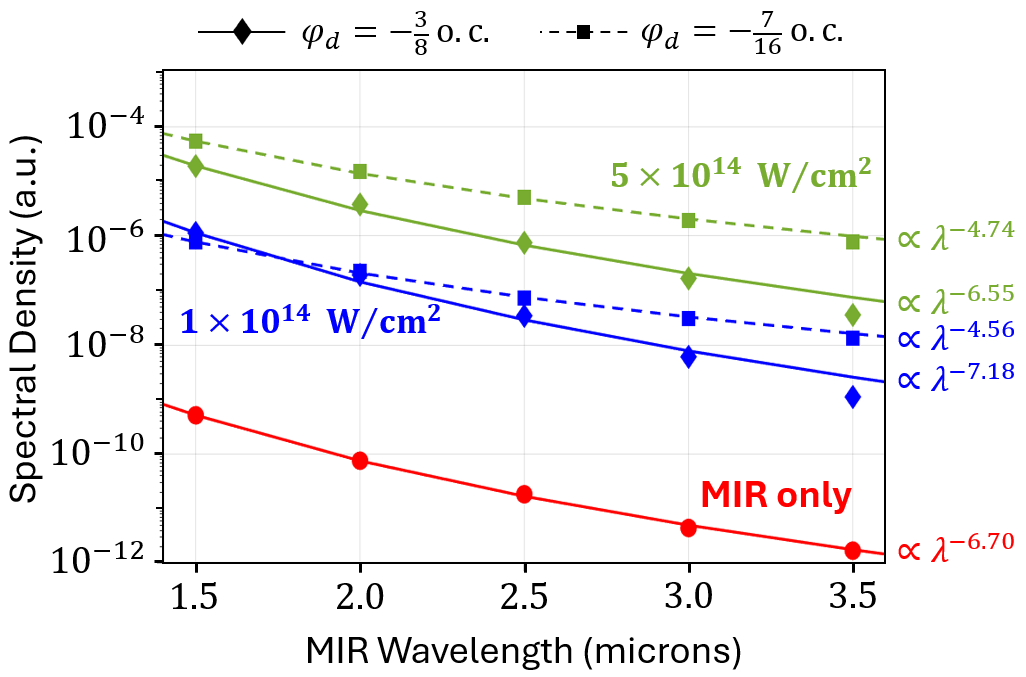}
  \caption{Efficiency scaling of the HHG plateau with differing wave lengths measuring from $2.22~U_p$ to $3.17~U_p$ where we compare two different phase delays $\phi_d=-7/16$ and $\phi_d=-3/8$ at two different intensity's $1\times10^{14}$, and $2\times10^{14}$ W/cm$^2$ to that of the MIR only.}
  \label{fig:3}
\end{figure}

Finally, we examine the effect of the UV pulse on the wavelength scaling of the single-atom HHG conversion efficiency. In the one-color (MIR-only) case, the efficiency was originally predicted to scale as $\lambda^{-3}$~\cite{lewenstein1994, tate2007}, while later studies reported steeper scalings of $\lambda^{-5}$ to $\lambda^{-6}$~\cite{shiner2009, yavuz2012, austin2012, frolov2015}. We evaluate the efficiency by integrating the HHG yield from the first plateau cutoff ($2.22U_{p}+I_{p}$) to the second plateau cutoff ($3.16U_{p}+I_{p}$), normalized by $U_{p}$. As shown in Fig.~\ref{fig:3}, the MIR-only case (red curve) follows a $\lambda^{-6.70}$ scaling, reflecting our definition of efficiency restricted to the higher part of the spectrum. With the addition of a UV pulse with an intensity of $10^{14}$~W/cm$^{2}$ (blue curves), we record the MIR-wavelength-dependent efficiency for two delays, $\phi_{d}=-3/8$ and $\phi_{d}=-7/16$ optical cycles.

For $\phi_{d}=-3/8$, the UV pulse launches predominantly short trajectories in the low-energy region of the HHG spectrum. As a result, the wavelength scaling ($\lambda^{-7.18}$) remains close to the MIR-only case. As $\phi_{d}$ approaches $-0.5$~o.c.\ (coinciding with the side peak of the MIR pulse), the UV pulse enhances both the higher-energy part of the spectrum and the contribution from long trajectories, as can be seen from the comparison between Fig.~\ref{fig:2}(c) and Fig.~\ref{fig:2}(d). This selective enhancement of the upper part of the HHG spectrum becomes more pronounced with increasing MIR wavelength, resulting in a more favorable wavelength scaling of $\lambda^{-4.56}$. The scaling improves over the MIR-only case because longer MIR wavelengths improve the selection of return trajectories only, while at shorter wavelengths, a larger fraction of these trajectories would ionize without recombination.

To confirm this interpretation, we also estimated the efficiency over a broader frequency range starting from the UV-only cutoff~\cite{supplemental}. As expected, both delays now yield similar scaling, more favorable than the MIR-only case, with further improvement at higher UV intensity. This improvement arises from the reduced ionization gate entering the tunneling regime and the refined selection of returning trajectories at longer MIR wavelengths. Consistent with the model of Lewenstein \textit{et al.}~\cite{lewenstein1994}, the inescapable $\lambda^{-3}$ scaling from wave-packet spreading remains, yet gating with a few-cycle UV pulse brings the efficiency closer to this ideal limit.

In conclusion, we have shown that a few-cycle UV pulse, which can be generated via RDW emission in gas-filled hollow-core fibers, provides control over the harmonic plateau by gating ionization within the MIR pulse. Even at modest UV intensities, the wavelength scaling of the HHG conversion efficiency improves significantly. Increasing the UV intensity and tuning the phase delay leads to a strong reduction in the attosecond pulse duration, reflecting the transition from multi-photon to tunneling ionization. Favoring short-trajectory contributions, as in earlier studies with attosecond pulse trains~\cite{schafer2004}, makes this two-color scheme robust against macroscopic phase matching. We expect these results to motivate further experimental and theoretical efforts toward optimizing this approach for attosecond science.


\bibliographystyle{apsrev}
\bibliography{bibliography.bib}

@article{piper2025,
  title = {Attosecond Clocking and Control of Strong Field Quantum Trajectories},
  author = {Piper, Andrew J. and Liu, Qiaoyi and Camacho Garibay, Abraham and Kiesewetter, Dietrich and Leshchenko, Vyacheslav and B\ae{}kh\o{}j, Jens E. and Agostini, Pierre and Schafer, Kenneth J. and DiMauro, Louis F. and Tang, Yaguo},
  journal = {Phys. Rev. Lett.},
  volume = {134},
  issue = {7},
  pages = {073201},
  numpages = {6},
  year = {2025},
  month = {2},
  publisher = {American Physical Society},
  doi = {10.1103/PhysRevLett.134.073201},
  url = {https://link.aps.org/doi/10.1103/PhysRevLett.134.073201}
}

@article{ammosov1986,
  author    = {M. V. Ammosov and N. B. Delone and V. P. Krainov},
  title     = {Tunnel ionization of complex atoms and atomic ions in a varying electromagnetic field},
  journal   = {Sov. Phys. JETP},
  volume    = {64},
  number    = {6},
  pages     = {1191--1194},
  year      = {1986}
}

@article{chen2014,
  author = {Ming-Chang Chen  and Christopher Mancuso  and Carlos Hernández-García  and Franklin Dollar  and Ben Galloway  and Dimitar Popmintchev  and Pei-Chi Huang  and Barry Walker  and Luis Plaja  and Agnieszka A. Jaroń-Becker  and Andreas Becker  and Margaret M. Murnane  and Henry C. Kapteyn  and Tenio Popmintchev },
  title = {Generation of bright isolated attosecond soft X-ray pulses driven by multicycle midinfrared lasers},
  journal = {Proceedings of the National Academy of Sciences},
  volume = {111},
  number = {23},
  pages = {E2361-E2367},
  year = {2014},
  doi = {10.1073/pnas.1407421111},
  URL = {https://www.pnas.org/doi/abs/10.1073/pnas.1407421111}
}

@article{biegert2021,
  author = {Belsa, B. and Amini, K. and Liu, X. and Sanchez, A. and Steinle, T. and Steinmetzer, J. and Le, A. T. and Moshammer, R. and Pfeifer, T. and Ullrich, J. and Moszynski, R. and Lin, C. D. and Gräfe, S. and Biegert, J.},
  title = {Laser-induced electron diffraction of the ultrafast umbrella motion in ammonia},
  journal = {Structural Dynamics},
  volume = {8},
  number = {1},
  pages = {014301},
  year = {2021},
  month = {01},
  issn = {2329-7778},
  doi = {10.1063/4.0000046},
  url = {https://doi.org/10.1063/4.0000046}
}

@article{xu2012,
  title = {Laser-Induced Electron Diffraction for Probing Rare Gas Atoms},
  author = {Xu, Junliang and Blaga, Cosmin I. and DiChiara, Anthony D. and Sistrunk, Emily and Zhang, Kaikai and Chen, Zhangjin and Le, Anh-Thu and Morishita, Toru and Lin, C. D. and Agostini, Pierre and DiMauro, Louis F.},
  journal = {Phys. Rev. Lett.},
  volume = {109},
  issue = {23},
  pages = {233002},
  numpages = {5},
  year = {2012},
  month = {12},
  publisher = {American Physical Society},
  doi = {10.1103/PhysRevLett.109.233002},
  url = {https://link.aps.org/doi/10.1103/PhysRevLett.109.233002}
}

@article{degiovannini2023,
  doi = {10.1088/1361-6455/acb872},
  url = {https://dx.doi.org/10.1088/1361-6455/acb872},
  year = {2023},
  month = {2},
  publisher = {IOP Publishing},
  volume = {56},
  number = {5},
  pages = {054002},
  author = {De Giovannini, Umberto and Küpper, Jochen and Trabattoni, Andrea},
  title = {New perspectives in time-resolved laser-induced electron diffraction},
  journal = {Journal of Physics B: Atomic, Molecular and Optical Physics}
}

@article{freeman1978,
  title = {Observation of Electric-Field-Induced Resonances above the Ionization Limit in a One-Electron Atom},
  author = {Freeman, R. R. and Economou, N. P. and Bjorklund, G. C. and Lu, K. T.},
  journal = {Phys. Rev. Lett.},
  volume = {41},
  issue = {21},
  pages = {1463--1467},
  numpages = {0},
  year = {1978},
  month = {Nov},
  publisher = {American Physical Society},
  doi = {10.1103/PhysRevLett.41.1463},
  url = {https://link.aps.org/doi/10.1103/PhysRevLett.41.1463}
}

@article{ni2008,
  title = {Above-threshold ionization in a strong dc electric field},
  author = {Ni, Y. and Zamith, S. and L\'epine, F. and Martchenko, T. and Kling, M. and Ghafur, O. and Muller, H. G. and Berden, G. and Robicheaux, F. and Vrakking, M. J. J.},
  journal = {Phys. Rev. A},
  volume = {78},
  issue = {1},
  pages = {013413},
  numpages = {7},
  year = {2008},
  month = {Jul},
  publisher = {American Physical Society},
  doi = {10.1103/PhysRevA.78.013413},
  url = {https://link.aps.org/doi/10.1103/PhysRevA.78.013413}
}

@article{ishikawa2003,
  title = {Photoemission and Ionization of ${\mathrm{H}\mathrm{e}}^{+}$ under Simultaneous Irradiation of Fundamental Laser and High-Order Harmonic Pulses},
  author = {Ishikawa, Kenichi},
  journal = {Phys. Rev. Lett.},
  volume = {91},
  issue = {4},
  pages = {043002},
  numpages = {4},
  year = {2003},
  month = {Jul},
  publisher = {American Physical Society},
  doi = {10.1103/PhysRevLett.91.043002},
  url = {https://link.aps.org/doi/10.1103/PhysRevLett.91.043002}
}

@article{johnson2019,
  author = {Johnson, Allan S.  and Avni, Timur  and Larsen, Esben W.  and Austin, Dane R.  and Marangos, Jon P. },
  title = {Attosecond soft X-ray high harmonic generation},
  journal = {Philosophical Transactions of the Royal Society A: Mathematical, Physical and Engineering Sciences},
  volume = {377},
  number = {2145},
  pages = {20170468},
  year = {2019},
  doi = {10.1098/rsta.2017.0468},
  URL = {https://royalsocietypublishing.org/doi/abs/10.1098/rsta.2017.0468}
}

@article{jin2018,
  author = {Cheng Jin and C. D. Lin},
  journal = {Photon. Res.},
  number = {5},
  pages = {434--442},
  publisher = {Optica Publishing Group},
  title = {Control of soft X-ray high harmonic spectrum by using two-color laser pulses},
  volume = {6},
  month = {5},
  year = {2018},
  url = {https://opg.optica.org/prj/abstract.cfm?URI=prj-6-5-434},
  doi = {10.1364/PRJ.6.000434}
}

@article{takahashi2004,
  author={Takahashi, E.J. and Nabekawa, Y. and Mashiko, H. and Hasegawa, H. and Suda, A. and Midorikawa, K.},
  journal={IEEE Journal of Selected Topics in Quantum Electronics}, 
  title={Generation of strong optical field in soft X-ray region by using high-order harmonics}, 
  year={2004},
  volume={10},
  number={6},
  pages={1315-1328},
  doi={10.1109/JSTQE.2004.838077}
}

@article{kroh2018,
  author = {Tobias Kroh and Cheng Jin and Peter Krogen and Philip D. Keathley and Anne-Laure Calendron and Jonathas P. Siqueira and Houkun Liang and Edilson L. Falc\~{a}o-Filho and C. D. Lin and Franz X. K\"{a}rtner and Kyung-Han Hong},
  journal = {Opt. Express},
  number = {13},
  pages = {16955--16969},
  publisher = {Optica Publishing Group},
  title = {Enhanced high-harmonic generation up to the soft X-ray region driven by mid-infrared pulses mixed with their third harmonic},
  volume = {26},
  month = {6},
  year = {2018},
  url = {https://opg.optica.org/oe/abstract.cfm?URI=oe-26-13-16955},
  doi = {10.1364/OE.26.016955}
}

@article{kim2004,
  author={Kim, I. J. and Kim, H. T. and Kim, C. M. and Park, J. J. and Lee, Y. S. and Hong, K.-H. and Nam, C. H.},
  title={Efficient high-order harmonic generation in a two-color laser field},
  journal={Applied Physics B},
  year={2004},
  month={5},
  day={01},
  volume={78},
  number={7},
  pages={859-861},
  doi={10.1007/s00340-004-1453-2},
  url={https://link.springer.com/article/10.1007/s00340-004-1453-2}
}

@article{pellegrini2016,
  title = {The physics of x-ray free-electron lasers},
  author = {Pellegrini, C. and Marinelli, A. and Reiche, S.},
  journal = {Rev. Mod. Phys.},
  volume = {88},
  issue = {1},
  pages = {015006},
  numpages = {55},
  year = {2016},
  month = {3},
  publisher = {American Physical Society},
  doi = {10.1103/RevModPhys.88.015006},
  url = {https://link.aps.org/doi/10.1103/RevModPhys.88.015006}
}

@article{ratner2019,
  title = {Pump-Probe Ghost Imaging with SASE FELs},
  author = {Ratner, D. and Cryan, J. P. and Lane, T. J. and Li, S. and Stupakov, G.},
  journal = {Phys. Rev. X},
  volume = {9},
  issue = {1},
  pages = {011045},
  year = {2019},
  month = {Mar},
  publisher = {American Physical Society},
  doi = {10.1103/PhysRevX.9.011045},
  url = {https://link.aps.org/doi/10.1103/PhysRevX.9.011045}
}

@article{duris2020,
  author={Duris, Joseph and Li, Siqi and Driver, Taran and Champenois, Elio G. and MacArthur, James P. and Lutman, Alberto A. and Zhang, Zhen and Rosenberger, Philipp and Aldrich, Jeff W. and Coffee, Ryan and others},
  title={Tunable isolated attosecond X-ray pulses with gigawatt peak power from a free-electron laser},
  journal={Nature Photonics},
  year={2020},
  month={1},
  volume={14},
  number={1},
  pages={30-36},
  issn={1749-4893},
  doi={10.1038/s41566-019-0549-5},
  url={https://doi.org/10.1038/s41566-019-0549-5}
}

@article{paul2001,
  author = {P. M. Paul and E. S. Toma and P. Breger and G. Mullot  and F. Aug\'{e} and P. Balcou and H. G. Muller and P. Agostini },
  title = {Observation of a Train of Attosecond Pulses from High Harmonic Generation},
  journal = {Science},
  volume = {292},
  number = {5522},
  pages = {1689-1692},
  year = {2001},
  doi = {10.1126/science.1059413},
  URL = {https://www.science.org/doi/abs/10.1126/science.1059413},
}

@article{ishikawa2007,
  title = {Single-attosecond pulse generation using a seed harmonic pulse train},
  author = {Ishikawa, Kenichi L. and Takahashi, Eiji J. and Midorikawa, Katsumi},
  journal = {Phys. Rev. A},
  volume = {75},
  issue = {2},
  pages = {021801},
  numpages = {4},
  year = {2007},
  month = {Feb},
  publisher = {American Physical Society},
  doi = {10.1103/PhysRevA.75.021801},
  url = {https://link.aps.org/doi/10.1103/PhysRevA.75.021801}
}

@article{travers2019,
  title = {High-energy pulse self-compression and ultraviolet generation through soliton dynamics in hollow capillary fibres},
  author = {John C. Travers and Teodora F. Grigorova and Christian Brahms and Federico Belli},
  journal = {Nat. Photonics},
  volume = {13},
  pages = {547–554},
  year = {2019},
  publisher = {Nature Publishing Group},
  doi = {10.1038/s41566-019-0416-4},
  url = {https://www.nature.com/articles/s41566-019-0416-4}
}

@article{wang2025,
  title = {Probing Electronic Coherence between Core-Level Vacancies at Different Atomic Sites},
  author = {Wang, Jun and Driver, Taran and Franz, Paris L. and Koloren\v{c}, P\v{r}emysl and Thierstein, Emily and Robles, River R. and Isele, Erik and Guo, Zhaoheng and Cesar, David and Alexander, Oliver and others},
  journal = {Phys. Rev. X},
  volume = {15},
  issue = {1},
  pages = {011008},
  year = {2025},
  month = {Jan},
  publisher = {American Physical Society},
  doi = {10.1103/PhysRevX.15.011008},
  url = {https://link.aps.org/doi/10.1103/PhysRevX.15.011008}
}

@article{huang2021,
  title = {Features and futures of X-ray free-electron lasers},
  journal = {The Innovation},
  volume = {2},
  number = {2},
  pages = {100097},
  year = {2021},
  issn = {2666-6758},
  doi = {https://doi.org/10.1016/j.xinn.2021.100097},
  url = {https://www.sciencedirect.com/science/article/pii/S2666675821000229},
  author = {Nanshun Huang and Haixiao Deng and Bo Liu and Dong Wang and Zhentang Zhao}
}

@article{shiner2009,
  title = {Wavelength Scaling of High Harmonic Generation Efficiency},
  author = {Shiner, A. D. and Trallero-Herrero, C. and Kajumba, N. and Bandulet, H.-C. and Comtois, D. and L\'egar\'e, F. and Gigu\`ere, M. and Kieffer, J-C. and Corkum, P. B. and Villeneuve, D. M.},
  journal = {Phys. Rev. Lett.},
  volume = {103},
  issue = {7},
  pages = {073902},
  numpages = {4},
  year = {2009},
  month = {8},
  publisher = {American Physical Society},
  doi = {10.1103/PhysRevLett.103.073902},
  url = {https://link.aps.org/doi/10.1103/PhysRevLett.103.073902}
}

@article{austin2012,
  title = {Strong-field approximation for the wavelength scaling of high-harmonic generation},
  author = {Austin, Dane R. and Biegert, Jens},
  journal = {Phys. Rev. A},
  volume = {86},
  issue = {2},
  pages = {023813},
  numpages = {7},
  year = {2012},
  month = {8},
  publisher = {American Physical Society},
  doi = {10.1103/PhysRevA.86.023813},
  url = {https://link.aps.org/doi/10.1103/PhysRevA.86.023813}
}

@article{frolov2015,
  title = {Scaling laws for high-order-harmonic generation with midinfrared laser pulses},
  author = {Frolov, M. V. and Manakov, N. L. and Xiong, Wei-Hao and Peng, Liang-You and Burgd\"orfer, J. and Starace, Anthony F.},
  journal = {Phys. Rev. A},
  volume = {92},
  issue = {2},
  pages = {023409},
  numpages = {8},
  year = {2015},
  month = {8},
  publisher = {American Physical Society},
  doi = {10.1103/PhysRevA.92.023409},
  url = {https://link.aps.org/doi/10.1103/PhysRevA.92.023409}
}

@article{yavuz2012,
  title = {Wavelength scaling of high-order-harmonic-generation efficiency by few-cycle laser pulses: Influence of carrier-envelope phase},
  author = {Yavuz, I. and Altun, Z. and Topcu, T.},
  journal = {Phys. Rev. A},
  volume = {86},
  issue = {4},
  pages = {043836},
  numpages = {4},
  year = {2012},
  month = {10},
  publisher = {American Physical Society},
  doi = {10.1103/PhysRevA.86.043836},
  url = {https://link.aps.org/doi/10.1103/PhysRevA.86.043836}
}

@article{lewenstein1994,
  title = {Theory of high-harmonic generation by low-frequency laser fields},
  author = {Lewenstein, M. and Balcou, Ph. and Ivanov, M. Yu. and L'Huillier, Anne and Corkum, P. B.},
  journal = {Phys. Rev. A},
  volume = {49},
  issue = {3},
  pages = {2117--2132},
  year = {1994},
  month = {3},
  publisher = {American Physical Society},
  doi = {10.1103/PhysRevA.49.2117},
  url = {https://link.aps.org/doi/10.1103/PhysRevA.49.2117}
}

@article{schafer2004,
  title = {Strong Field Quantum Path Control Using Attosecond Pulse Trains},
  author = {Schafer, Kenneth J. and Gaarde, Mette B. and Heinrich, Arne and Biegert, Jens and Keller, Ursula},
  journal = {Phys. Rev. Lett.},
  volume = {92},    
  issue = {2},
  pages = {023003},
  numpages = {4},
  year = {2004},
  month = {1},
  publisher = {American Physical Society},
  doi = {10.1103/PhysRevLett.92.023003},
  url = {https://link.aps.org/doi/10.1103/PhysRevLett.92.023003}
}

@article{biegert2006,
  author = {J. Biegert and A. Heinrich and C. P. Hauri and W. Kornelis and P. Schlup and M. P. Anscombe and M. B. Gaarde and K. J. Schafer and U. Keller},
  title = {Control of high-order harmonic emission using attosecond pulse trains},
  journal = {Journal of Modern Optics},
  volume = {53},
  number = {1-2},
  pages = {87--96},
  year = {2006},
  publisher = {Taylor \& Francis},
  doi = {10.1080/09500340500167669},
  url={https://doi.org/10.1080/09500340500167669}
}

@article{gademann2011,
  doi = {10.1088/1367-2630/13/3/033002},
  url = {https://dx.doi.org/10.1088/1367-2630/13/3/033002},
  year = {2011},
  month = {3},
  publisher = {IOP Publishing},
  volume = {13},
  number = {3},
  pages = {033002},
  author = {Gademann, G and Kelkensberg, F and Siu, W K and Johnsson, P and Gaarde, M B and Schafer, K J and Vrakking, M J J},
  title = {Attosecond control of electron–ion recollision in high harmonic generation},
  journal = {New Journal of Physics}
}

@article{heldt2023,
  title = {Attosecond Real-Time Observation of Recolliding Electron Trajectories in Helium at Low Laser Intensities},
  author = {Heldt, Tobias and Dubois, Jonathan and Birk, Paul and Borisova, Gergana D. and Lando, Gabriel M. and Ott, Christian and Pfeifer, Thomas},
  journal = {Phys. Rev. Lett.},
  volume = {130},
  issue = {18},
  pages = {183201},
  numpages = {6},
  year = {2023},
  month = {5},
  publisher = {American Physical Society},
  doi = {10.1103/PhysRevLett.130.183201},
  url = {https://link.aps.org/doi/10.1103/PhysRevLett.130.183201}
}

@article{mauritsson2006,
  title = {Attosecond Pulse Trains Generated Using Two Color Laser Fields},
  author = {Mauritsson, J. and Johnsson, P. and Gustafsson, E. and L'Huillier, A. and Schafer, K. J. and Gaarde, M. B.},
  journal = {Phys. Rev. Lett.},
  volume = {97},
  issue = {1},
  pages = {013001},
  numpages = {4},
  year = {2006},
  month = {7},
  publisher = {American Physical Society},
  doi = {10.1103/PhysRevLett.97.013001},
  url = {https://link.aps.org/doi/10.1103/PhysRevLett.97.013001}
}

@article{watanabe1994,
  title = {Two-Color Phase Control in Tunneling Ionization and Harmonic Generation by a Strong Laser Field and Its Third Harmonic},
  author = {Watanabe, S. and Kondo, K. and Nabekawa, Y. and Sagisaka, A. and Kobayashi, Y.},
  journal = {Phys. Rev. Lett.},
  volume = {73},
  issue = {20},
  pages = {2692--2695},
  year = {1994},
  month = {11},
  publisher = {American Physical Society},
  doi = {10.1103/PhysRevLett.73.2692},
  url = {https://link.aps.org/doi/10.1103/PhysRevLett.73.2692}
}

@article{perry1993,
  title = {High-order harmonic emission from mixed fields},
  author = {Perry, Michael D. and Crane, John K.},
  journal = {Phys. Rev. A},
  volume = {48},
  issue = {6},
  pages = {R4051--R4054},
  year = {1993},
  month = {Dec},
  publisher = {American Physical Society},
  doi = {10.1103/PhysRevA.48.R4051},
  url = {https://link.aps.org/doi/10.1103/PhysRevA.48.R4051}
}

@article{schafer1993,
  title = {Above threshold ionization beyond the high harmonic cutoff},
  author = {Schafer, K. J. and Yang, Baorui and DiMauro, L. F. and Kulander, K. C.},
  journal = {Phys. Rev. Lett.},
  volume = {70},
  issue = {11},
  pages = {1599--1602},
  year = {1993},
  month = {3},
  publisher = {American Physical Society},
  doi = {10.1103/PhysRevLett.70.1599},
  url = {https://link.aps.org/doi/10.1103/PhysRevLett.70.1599}
}

@book{chang2016,
  title={Fundamentals of Attosecond Optics},
  author={Chang, Z.},
  isbn={9781420089387},
  lccn={2011016703},
  url={https://books.google.com/books?id=J5HLBQAAQBAJ},
  year={2016},
  publisher={CRC Press}
}

@article{antoine1996,
  title = {Attosecond Pulse Trains Using High-Order Harmonics},
  author = {Antoine, Philippe and L'Huillier, Anne and Lewenstein, Maciej},
  journal = {Phys. Rev. Lett.},
  volume = {77},
  issue = {7},
  pages = {1234--1237},
  numpages = {0},
  year = {1996},
  month = {8},
  publisher = {American Physical Society},
  doi = {10.1103/PhysRevLett.77.1234},
  url = {https://link.aps.org/doi/10.1103/PhysRevLett.77.1234}
}

@article{hentschel2001,
  title = {Attosecond metrology},
  author = {Hentschel, M. and Kienberger, R. and Spielmann, Ch. and Reider, G. A. and Milosevic, N. and Brabec, T. and Corkum, P. and Heinzmann, U. and Drescher, M. and Krausz, F.},
  journal = {Nature},
  volume = {414},
  issue = {6863},
  pages = {509-513},
  numpages = {4},
  year = {2001},
  month = {11},
  publisher = {Springer Nature},
  doi = {10.1038/351070004},
  url = {https://doi.org/10.1038/35107000}
}

@article{kling2008,
  author = {Kling, M. F. and Vrakking, M. J. J.},
  title = {Attosecond Electron Dynamics},
  journal = {Annual Review of Physical Chemistry},
  volume = {59},
  number = {1},
  pages = {463-492},
  year = {2008},
  doi = {10.1146/annurev.physchem.59.032607.093532},
  URL = {https://doi.org/10.1146/annurev.physchem.59.032607.093532}
}

@article{krausz2009,
  title = {Attosecond physics},
  author = {Krausz, F. and Ivanov, M.},
  journal = {Rev. Mod. Phys.},
  volume = {81},
  issue = {1},
  pages = {163--234},
  numpages = {0},
  year = {2009},
  month = {2},
  publisher = {American Physical Society},
  doi = {10.1103/RevModPhys.81.163},
  url = {https://link.aps.org/doi/10.1103/RevModPhys.81.163}
}

@article{altucci2010,
  author = {Altucci, C. and Velotta, R. and Marangos, J. P.},
  year = {2010},
  month = {7},
  pages = {916-952},
  title = {Ultra-fast dynamic imaging: An overview of current techniques, their capabilities and future prospects},
  volume = {57},
  journal = {Journal of Modern Optics},
  doi = {10.1080/09500340.2010.493621}
}

@article{corkum1993,
  title = {Plasma perspective on strong field multiphoton ionization},
  author = {Corkum, P. B.},
  journal = {Phys. Rev. Lett.},
  volume = {71},
  issue = {13},
  pages = {1994--1997},
  numpages = {0},
  year = {1993},
  month = {9},
  publisher = {American Physical Society},
  doi = {10.1103/PhysRevLett.71.1994},
  url = {https://link.aps.org/doi/10.1103/PhysRevLett.71.1994}
}

@article{ramasesha2016,
  author = {Ramasesha, Krupa and Leone, Stephen R. and Neumark, Daniel M.},
  title = {Real-Time Probing of Electron Dynamics Using Attosecond Time-Resolved Spectroscopy}, 
  journal= {Annual Review of Physical Chemistry},
  year = {2016},
  volume = {67},
  pages = {41-63},
  doi = {https://doi.org/10.1146/annurev-physchem-040215-112025},
  url = {https://www.annualreviews.org/content/journals/10.1146/annurev-physchem-040215-112025},
  publisher = {Annual Reviews}
}

@article{ren2018,
  doi = {10.1088/2040-8986/aaa394},
  url = {https://dx.doi.org/10.1088/2040-8986/aaa394},
  year = {2018},
  month = {1},
  publisher = {IOP Publishing},
  volume = {20},
  number = {2},
  pages = {023001},
  author = {Ren, Xiaoming and Li, Jie and Yin, Yanchun and Zhao, Kun and Chew, Andrew and Wang, Yang and Hu, Shuyuan and Cheng, Yan and Cunningham, Eric and Wu, Yi and Chini, Michael and Chang, Zenghu},
  title = {Attosecond light sources in the water window},
  journal = {Journal of Optics}
}

@article{trabattoni2019,
  author = {Trabattoni, A.  and Galli, M.  and Lara-Astiaso, M.  and Palacios, A.  and Greenwood, J.  and Tavernelli, I.  and Decleva, P.  and Nisoli, M.  and Martín, F.  and Calegari, F. },
  title = {Charge migration in photo-ionized aromatic amino acids},
  journal = {Philosophical Transactions of the Royal Society A: Mathematical, Physical and Engineering Sciences},
  volume = {377},
  number = {2145},
  pages = {20170472},
  year = {2019},
  doi = {10.1098/rsta.2017.0472},
  URL = {https://royalsocietypublishing.org/doi/abs/10.1098/rsta.2017.0472}
}

@article{jordan2020,
  author = {Inga Jordan and Martin Huppert  and Dominik Rattenbacher  and Michael Peper  and Denis Jelovina  and Conaill Perry  and Aaron von Conta  and Axel Schild  and Hans Jakob Wörner },
  title = {Attosecond spectroscopy of liquid water},
  journal = {Science},
  volume = {369},
  number = {6506},
  pages = {974-979},
  year = {2020},
  doi = {10.1126/science.abb0979},
  URL = {https://www.science.org/doi/abs/10.1126/science.abb0979}
}

@article{zinchenko2023,
  author={Zinchenko, Kristina S. and Ardana-Lamas, Fernando and Lanfaloni, Valentina Utrio and Luu, Tran Trung and Pertot, Yoann and Huppert, Martin and W{\"o}rner, Hans Jakob},
  title={Apparatus for attosecond transient-absorption spectroscopy in the water-window soft-X-ray region},
  journal={Scientific Reports},
  year={2023},
  month={2},
  day={21},
  volume={13},
  number={1},
  pages={3059},
  doi={10.1038/s41598-023-29089-8},
  url={https://doi.org/10.1038/s41598-023-29089-8}
}

@incollection{mauger2025,
  title = {Attosecond charge migration in organic molecules: Initiating and probing localized electron holes},
  editor = {Louis F. Dimauro and Hélène Perrin and Susanne Yelin},
  series = {Advances In Atomic, Molecular, and Optical Physics},
  publisher = {Academic Press},
  volume = {74},
  pages = {1-46},
  year = {2025},
  issn = {1049-250X},
  doi = {https://doi.org/10.1016/bs.aamop.2025.04.001},
  url = {https://www.sciencedirect.com/science/article/pii/S1049250X25000011},
  author = {François Mauger and Sucharita Giri and Aderonke S. Folorunso and Kyle A. Hamer and Denawakage D. Jayasinghe and Kenneth Lopata and Kenneth J. Schafer and Mette B. Gaarde}
}

@article{grell2023,
  title = {Effect of the shot-to-shot variation on charge migration induced by sub-fs x-ray free-electron laser pulses},
  author = {Grell, G. and Guo, Z. and Driver, T. and Decleva, P. and Pl\'esiat, E. and Pic\'on, A. and Gonz\'alez-V\'azquez, J. and Walter, P. and Marangos, J. P. and Cryan, J. P. and others},
  journal = {Phys. Rev. Res.},
  volume = {5},
  issue = {2},
  pages = {023092},
  numpages = {12},
  year = {2023},
  month = {5},
  publisher = {American Physical Society},
  doi = {10.1103/PhysRevResearch.5.023092},
  url = {https://link.aps.org/doi/10.1103/PhysRevResearch.5.023092}
}

@misc{supplemental,
  note = {See Supplemental Material at [URL will be inserted by publisher] for additional details.}
}

\newpage


\section*{Supplementary material}

# Supplemental Material for "Generation of Bright and Controllable Isolated Attosecond X-Ray Pulses from Synchronized Mid-Infrared and Ultra-short Ultraviolet Laser Fields"

Davis Robinson,[1, *] Kyle A. Hamer,[1] Chelsea Kincaid,[2] Michael Chini,[2] and Nicolas Douguet[1, †]

[1]*Department of Physics, University of Central Florida, Orlando, FL 32816, USA*
[2]*The Ohio State University, Columbus, OH 43210, USA*
(Dated: November 15, 2025)

## I. THEORETICAL MODELS

### A. Single-active electron (SAE) model

The TDSE was solved within the SAE approximation using the code described in detail in Refs. [1, 2]. Specifically, we solved the nonrelativistic TDSE in the dipole approximation and in the length gauge, given by

$$i\partial_t|\psi(t)\rangle = [\hat{H}_0 + \hat{H}_{\text{int}}(t)]|\psi(t)\rangle \ . \tag{1}$$

In the above equation, $\hat{H}_0 = \hat{T} + \hat{V}$ is the field-free Hamiltonian, where $\hat{T}$ and $\hat{V}$ are the kinetic- and potential-energy operators, respectively. The light–matter interaction is taken in the length gauge, $\hat{H}_{\text{int}}(t) = \boldsymbol{r} \cdot \boldsymbol{E}(t)$, with $\boldsymbol{r}$ the electron position vector and $\boldsymbol{E}(t)$ the electric field.

We employ a bichromatic linearly-polarized electric field along the $\hat{x}$ axis of the form:

$$\boldsymbol{E}(t) = [E_{\text{MIR}}(t) + E_{\text{UV}}(t - \tau)]\,\hat{x} \ , \tag{2}$$

where the MIR and UV electric fields are given by

$$E_{\text{MIR}}(t) = f_{\text{MIR}}(t)\cos(\omega_{\text{MIR}}t); \quad E_{\text{UV}}(t) = f_{\text{UV}}(t)\cos(\omega_{\text{UV}}t + \phi_{\text{UV}}) \ . \tag{3}$$

In the above equations, $\omega_{\text{MIR}}$ and $\omega_{\text{UV}}$ are the MIR and UV frequencies, respectively, $\phi_{\text{UV}}$ is the carrier-envelope phase (CEP) of the UV pulse, and $\tau$ is the time delay between the MIR and UV pulses. For the main calculations presented in this work, the CEP was set to $\phi_{\text{UV}} = 0$. In the main manuscript, we express the delay in units of optical cycles (o.c.) of the MIR pulse for convenience, defining $\phi_d = \tau/T_{\text{MIR}}$, where $T_{\text{MIR}}$ denotes the MIR period.

For both the MIR and UV pulses, we employ sine-square envelopes of the form

$$f_{\text{MIR}}(t) = E_{\text{MIR}}\sin^2\left(\frac{\omega_{\text{MIR}}t}{N_{\text{MIR}}}\right); \qquad f_{\text{UV}}(t) = E_{\text{UV}}\sin^2\left(\frac{\omega_{\text{UV}}t}{N_{\text{UV}}}\right), \tag{4}$$

where $E_{\text{MIR}}$ and $E_{\text{UV}}$ denote the peak electric fields of the MIR and UV pulses, respectively, while $N_{\text{MIR}}$ and $N_{\text{UV}}$ are the corresponding numbers of optical cycles.

We describe helium using the model potential of Tong and Lin [3], which has been shown to provide excellent agreement for HHG spectra computed with long-wavelength pulses when compared with full two-electron calculations performed using the $R$-matrix with time dependence [4]. Using this model potential, we obtain the helium ground state $|\psi_0\rangle$ by diagonalizing the field-free Hamiltonian, such that $\hat{H}_0|\psi_0\rangle = -I_p|\psi_0\rangle$, with an ionization potential $I_p = 0.901$ a.u., in excellent agreement with the experimental value.

All calculations employ a uniform radial grid with step size $h = 0.05$ a.u. The partial-wave expansion of the wavefunction $|\psi(t)\rangle$ includes angular momenta up to $\ell_{\text{max}} = 1200$ for the most demanding cases (MIR wavelength $\lambda_{\text{MIR}} = 3500$ nm). The radial box size is set to $R_{\text{max}} = 2.5\,\alpha_0$, where the quiver amplitude is $\alpha_0 = E_{\text{MIR}}/\omega_{\text{MIR}}^2$, ensuring that both short- and long-trajectory dynamics are contained. To suppress spurious reflections, a quartic complex absorbing potential is applied starting 50 a.u. before the end of the grid.

We propagate the TDSE from the helium ground state, $|\psi(0)\rangle = |\psi_0\rangle$, using a second-order split-operator scheme. The field-free step is implemented with a Crank–Nicolson method, which yields a tridiagonal linear system solved by the Thomas algorithm. This combination provides unitary, second-order–accurate time evolution.

_____

[*] davis.robinson@ucf.edu
[†] nicolas.douguet@ucf.edu

The induced dipole moment in helium is calculated in the acceleration form of the dipole operator as

$$\boldsymbol{a}(t) = \langle \psi(t)|\nabla V|\psi(t)\rangle + \boldsymbol{E}(t) \ . \tag{5}$$

We then compute the Fourier transform of the dipole moment as

$$\tilde{\boldsymbol{a}}(\omega) = \int_{-\infty}^{\infty} W(t)e^{-i\omega t}\boldsymbol{a}(t)dt \ . \tag{6}$$

Here, $W(t)$ is a window function that equals unity during the pulse and then tapers with a sine-squared envelope over an additional half optical cycle after the end of the pulse, thereby suppressing ringing from residual excited-state populations and yielding a clean, low-noise spectrum.

Finally, the spectra density is computed as

$$S(\omega) = \frac{2}{3\pi c}|\tilde{\boldsymbol{a}}(\omega)|^2 \ . \tag{7}$$

For all calculations presented in this work, we verified spectral convergence with respect to the TDSE parameters (time step, radial step, and $\ell_{\max}$), and confirmed that spectra computed using the length or acceleration form of the dipole produce indistinguishable spectra.

## B. Comparison with the semi-classical model

To demonstrate control of the HHG signal via the delay between the MIR and UV pulses, we computed MIR+UV spectra on a fine grid of phase delays with step size $\Delta\phi_d = 0.01$ o.c., for $\phi_d \in [-0.5, 0.5]$. The MIR and UV intensities were fixed at $2 \times 10^{14}$ and $10^{14}$ W/cm², respectively.

Figure 1 shows results for a 2 $\mu$m MIR driver. Even at this relatively modest MIR wavelength, a 1.5 fs UV pulse enables single-cycle control of the plateau cutoff, whereas finer trajectory selection and sub-cycle control improve with increasing MIR wavelength. The spectra exhibit the expected variation of the plateau cutoff and reveal delay intervals in which electrons launched by the UV field do not recombine with the parent ion.

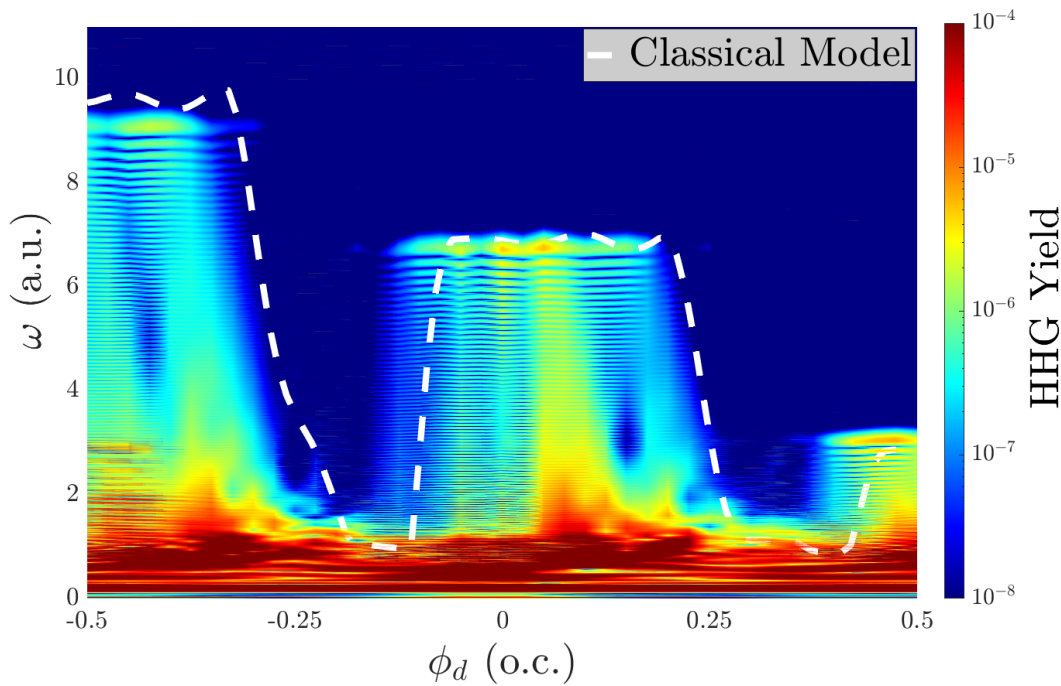

FIG. 1. Color map of the HHG spectra, obtained from TDSE simulations, driven by a 2 $\mu$m MIR pulse and a 244 nm UV pulse, shown as a function of the phase delay $\phi_d$. For comparison, the classical-model prediction of the maximum returning-electron (recombination) kinetic energy $K_{\max}(\phi_d) + I_p$ is overlaid (dashed white curve).

To check that the plateau structure match the expectations from the classical model, we solved Newton's equation for an electron released with zero velocity in a linearly-polarized electric field combining the MIR and delayed UV pulses. To take into account the temporal width of the UV pulse, we launch all trajectories belonging to the FWHM of the UV electric field amplitude and record the maximum kinetic electron energy at recombination, $K_{\max}(\phi_d)$, at a fixed phase delay $\phi_d$. As a result, we observe an overall very good agreement between the semi-classical and the TDSE results (see Fig. 1). The additional complex delay-dependent features in the HHG signal can be better analyzed by performing Gabor transforms of the HHG spectra (see main text).

### C. Strong-field approximation (SFA) calculations

We also carried out calculations within the strong-field approximation (SFA) to assess whether this simple model can reproduce the main features of the two-color scheme with widely disparate pulses, and to provide an efficient means of predicting outcomes across pulse parameters.

In the SFA, contributions from bound excited states are neglected and, after tunneling, the Coulomb interaction between the continuum electron and the parent ion is ignored. The dipole moment is computed under the assumption that no intermediate bound–bound couplings occur and that only the ground–continuum transition contributes [5, 6]. The time-dependent dipole in the Lewenstein model then reads

$$\boldsymbol{D}(t) = i \int_{-\infty}^{t} dt' \int d^3\boldsymbol{p} \, \boldsymbol{d}^*(\boldsymbol{p} + \boldsymbol{A}(t)) \exp[-iS(\boldsymbol{p}, t, t')] \, \boldsymbol{E}(t') \cdot \boldsymbol{d}(\boldsymbol{p} + \boldsymbol{A}(t')) + \text{c.c.} \tag{8}$$

In Eq. (8), the term $\boldsymbol{d}(\boldsymbol{p}) = \langle \boldsymbol{p} | \boldsymbol{r} | \psi_0 \rangle$ represents the bound-free transition dipole moment between the ground state $|\psi_0\rangle$ and the plane wave state, $|\boldsymbol{p}\rangle$, while the phase factor

$$S(\boldsymbol{p}, t, t') = \int_{t'}^{t} \left( [\boldsymbol{p} + A(t'')]^2 + I_p \right) dt'' \;, \tag{9}$$

is the modified classical action, where $I_p = 0.903$ a.u. is the ionization potential of helium.

We adopt a Gaussian model in which the initial state is a normalized Gaussian of width $\Delta$,

$$\langle \boldsymbol{r} | \psi_0 \rangle = \left( \pi \Delta^2 \right)^{-\frac{3}{4}} e^{-\boldsymbol{r}^2 / 2\Delta^2} \;, \tag{10}$$

leading to an analytical expression for the bound-free transition dipole moment of the form

$$d(\boldsymbol{p}) = i \left( \frac{\Delta^2}{\pi} \right)^{3/4} \Delta^2 \frac{\boldsymbol{p}}{\alpha} e^{-\Delta^2 \boldsymbol{p}^2 / 2} \;. \tag{11}$$

The Gaussian model is employed to show that, even at the simplest level of theory, the SFA reproduces the main features of the MIR+UV scheme, indicating its general applicability across different systems. In addition, the Gaussian model for the dipole matrix elements is free from singularities in the complex time plane and yields a closed-form expression for the dipole along the polarization axis, $x(t) = \boldsymbol{D}(t) \cdot \hat{x}$, as

$$\begin{aligned} x(t) = i\Delta^{-7} \int_{0}^{t} & 2C(t,t')^{3/2} E(t') \\ & \times \{ A(t)A(t') + C(t,t')[1 - D(t,t')(A(t) + A(t'))] + C^2(t,t'D^2(t,t')) \} \\ & \times \exp\left( -i[I_p(t-t') + B(t,t')] - \frac{[A^2(t) + A^2(t')]\Delta^2 - C(t,t')D^2(t,t')}{2} \right) dt' \;, \end{aligned} \tag{12}$$

where $\boldsymbol{A}(t) = -\int_{-\infty}^{t} \boldsymbol{E}(t')dt'$ is the vector potential, and the functions $B$, $C$, and $D$ are expressed as

$$B(t,t') = \frac{1}{2} \int_{t'}^{t} A^2(t'')dt'' \;; \tag{13}$$

$$C(t,t') = \frac{1}{2\Delta^2 + i(t-t')} \;; \tag{14}$$

$$D(t,t') = [A(t) + A(t')]\Delta^2 + i \int_{t'}^{t} A(t'')dt'' \;. \tag{15}$$

As in the TDSE calculations, we employ few-cycle MIR and UV pulses with a controllable delay $\phi_d$. The MIR driver has wavelength 3000 nm, consists of four optical cycles with a $\sin^2$ envelope, and has a peak intensity of $2 \times 10^{14}$ W/cm². For the UV pulse, we adopt a sine-fourth envelope with peak intensity $10^{14}$ W/cm² and FWHM $\sim 1.5$ fs, as it mitigates low-frequency spectral spikes in HHG driven by the UV pulse alone.

In the Gaussian model, the exponential falloff of the bound–continuum transition dipole with electron momentum requires choosing the width $\Delta$ such that $\Delta < \sqrt{2/\omega_c}$ [7], where $\omega_c$ is the harmonic cutoff frequency; this prevents an artificial drop of the HHG signal before the plateau cutoff. In all SFA calculations, we set $\Delta = 0.2$ a.u. As before, we apply a sine-square window function over an additional half optical cycle after the end of the MIR pulse to suppress background contributions. The HHG spectrum is computed from the dipole acceleration, obtained by taking the second time derivative of the dipole moment in Eq. (12).

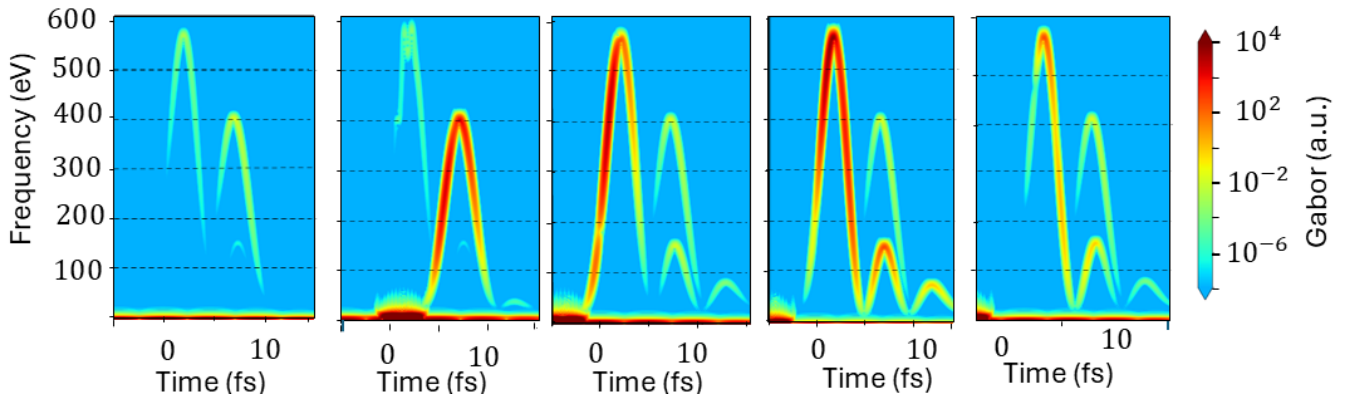

FIG. 2. Gabor transforms of the dipole acceleration for (a) MIR-only, and three different values of the phase delay $\phi_d$: (b) $-5/8$ o.c., (c) $-1/2$ o.c., and (d) $-3/8$ o.c..

.

The Gabor transforms computed within the SFA for the MIR-only case and for the MIR+UV field at selected phase delays are shown in Fig. 2. The SFA reproduces the main features of the TDSE results: enhancement of the HHG signal, selective control of short and long trajectories, and filtering of multiple-rescattering contributions. Moreover, the magnitude of the signal enhancement is in quantitative agreement with the TDSE calculations.

## II. INTENSITY SCALING AND TEMPORAL WIDTH OF THE ISOLATED ATTOSECOND PULSE

### A. Analysis of the attosecond peak intensity

Within the dipole approximation and in the length gauge, the SAE Hamiltonian for helium driven by a combined MIR+UV field reads

$$H(t) = -\frac{1}{2}\nabla^2 + V_{\mathrm{He}}(r) + \boldsymbol{r} \cdot \left[\boldsymbol{E}_{\mathrm{MIR}}(t) + \boldsymbol{E}_{\mathrm{UV}}(t)\right] , \tag{16}$$

For linearly polarized fields along $\hat{x}$, and assuming the MIR field is quasi-static over the UV pulse duration, we take $\boldsymbol{E}_{\mathrm{MIR}}(t) \simeq E_{\mathrm{MIR}}(\phi_d)\hat{x}$. Then, in the UV time window,

$$H(t) \simeq -\frac{1}{2}\nabla^2 + V_{\mathrm{eff}}(\mathbf{r}) - xE_{\mathrm{UV}}(t) , \qquad V_{\mathrm{eff}}(\mathbf{r}) = V_{\mathrm{He}}(r) - x\,E_{\mathrm{MIR}}(\phi_d) . \tag{17}$$

Along the positive $\hat{x}$ axis, we replace the helium SAE potential by its Coulomb tail (valid away from the core),

$$V_{\mathrm{eff}}(x) = -\frac{1}{x} - x\,|E_{\mathrm{MIR}}(\phi_d)| , \tag{18}$$

where we take $E_{\mathrm{MIR}}(\phi_d) < 0$ by convention. The effective potential has a barrier (saddle) at

$$x_b = |E_{\mathrm{MIR}}(\phi_d)|^{-1/2} , \qquad V_{\mathrm{eff}}(x_b) = -2\sqrt{|E_{\mathrm{MIR}}(\phi_d)|} . \tag{19}$$

Thus the barrier is lowered by $\Delta V = 2\sqrt{|E_{\text{MIR}}(\phi_d)|}$. Neglecting the small DC Stark shift of the ground state in the quasi-static MIR field, and assuming that ionization by the UV occurs in the perturbative multiphoton regime ($\gamma_{\text{UV}} \gg 1$), the ionization steps proceeds against an effective ionization potential $I_p^{\text{eff}} = I_p - \Delta V$, such that the minimum number of UV photons required to ionize is $N_p = \lceil I_p^{\text{eff}}/\omega_{\text{UV}} \rceil$. At $\phi_d = -3/8$, the quasi-static MIR field is $|E_{\text{MIR}}(\phi_d)| \simeq 0.057$ a.u., which lowers the barrier by $\Delta V = 2\sqrt{|E_{\text{MIR}}|} \simeq 0.48$ a.u., giving an effective ionization potential $I_p^{\text{eff}} \simeq 0.42$ a.u. The ratio $I_p^{\text{eff}}/\omega_{\text{UV}} \simeq 2.26$ places the process between two- and three-photon ionization taking into account the broad UV bandwidth and nonadiabatic effects of the MIR.

Our TDSE calculations at this phase, shown in Fig. 3, yield an ionization probability that follows a power law $P_{\text{ion}} \propto I_{\text{UV}}^{\alpha}$ with $\alpha \simeq 2.26$, fully consistent with the barrier suppression model. For this phase, most emitted trajectories can recombine, so the peak intensity of the generated attosecond pulse inherits a similar $I_{\text{UV}}^{\alpha}$ scaling.

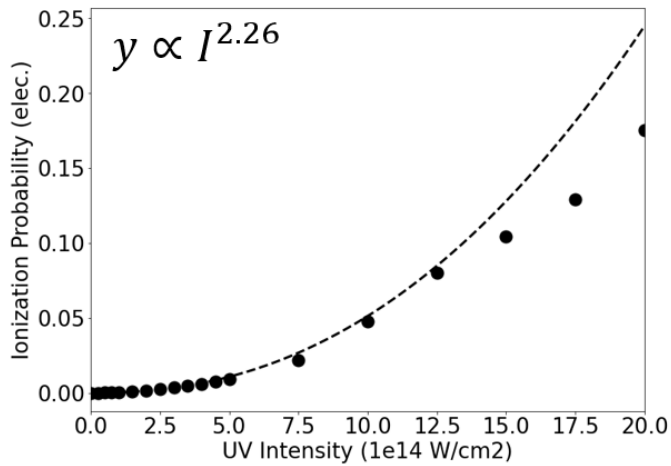

FIG. 3. Ionization probability computed with the TDSE at $\phi_d = -3/8$ as a function of the UV peak intensity. The trendline is obtained by fitting points below $I_{\text{UV}} = 3.5 \times 10^{14}$ W/cm$^2$ (the multiphoton limit for this particular UV wavelength).

Finally, we also tested this model by performing SFA calculations at three different MIR intensities, 1.5, 2.0, and $2.5 \times 10^{14}$ W/cm$^2$, for a fixed phase delay $\phi_d = -3/8$. As the MIR intensity increases, the barrier is lowered more efficiently, leading to a higher peak intensity of the attosecond pulse, consistent with the trend observed in Fig. 4. In the multiphoton regime of the UV pulse, the peak intensity follows a power law $I_{\text{UV}}^{\alpha_{\text{MIR}}}$, with an effective exponent $\alpha_{\text{MIR}}$ that decreases as the MIR intensity rises, reflecting the reduced number of photons required for ionization. The power-law exponent obtained from the SFA at a MIR intensity of $2 \times 10^{14}$ W/cm$^2$ from SFA is seen to be in good agreement with the TDSE results.

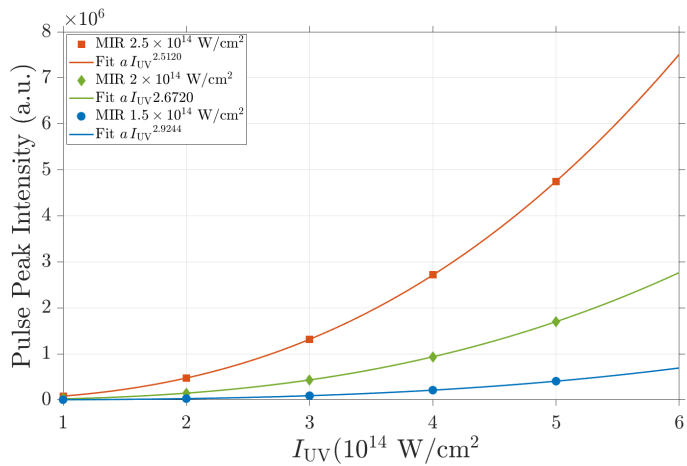

FIG. 4. Peak intensity of the attosecond pulse, calculated at three MIR intensities, as a function of the UV intensity. The trends for all MIR intensities follow a power law $I_{\text{UV}}^{\alpha_{\text{MIR}}}$.

## B. Analysis of the attosecond temporal width

We consider a UV pulse of the form

$$E(t) = E_0 f(t) \cos(\omega_{UV} t) \ , \tag{20}$$

where $E_0$ is the peak field amplitude and $0 \leq f(t) \leq 1$ is the pulse envelope with characteristic temporal width $\sigma_{UV}$. We also set the time origin at the envelope maximum, $f(0) = 1$.

In the perturbative UV multiphoton regime, treating the MIR field as quasi-static over the UV pulse and assuming a slow varying envelope over the UV period, $(\omega_{UV}\sigma_{UV} \gg 1)$, the cycle-averaged $N$-photon ionization rate is

$$W_N(t) \propto I_0^N f^N(t) \ , \tag{21}$$

where $I_0 = E_0^2/2$ is the peak intensity. Consequently, the temporal width of the rate is reduced; for a Gaussian envelope it is smaller than the UV envelope by exactly a factor $1/\sqrt{N}$ compared to the UV envelope. For a three-photon process, including barrier suppression, the FWHM of the ionization rate is about 866 as. At a phase delay favoring short trajectories ($\phi_d = -3/8$ o.c.), once the recombination time and the suppression of lower-frequency components are accounted for, this is consistent with generating a high-energy pulse with FWHM of about 1 fs.

For intermediate Keldysh parameters (nonadiabatic tunneling), the Yudin–Ivanov model [8] predicts that the instantaneous ionization rate narrows near the field maxima. As the system approaches the tunneling regime ($\gamma \simeq 1$), the Yudin–Ivanov rate reduces to the ADK form

$$W_{ADK}(t) = C|E(t)|^{-p} \exp\left(-\frac{2\kappa^3}{3|E(t)|}\right), \tag{22}$$

where $\kappa = (2I_p)^{1/2}$, $n^* = 1/\kappa$, $p = 2n^* - 1$, and $C$ is a target-dependent constant. In this regime, the instantaneous field $E(t)$ is set by the combined MIR and UV fields, and $I_p$ is the helium ionization potential. For high UV field, we can neglect the effect of the MIR field, and expand the field around the UV maximum, $E(t) \simeq E_0(1 - \omega_{UV}^2 t^2/2)$, to obtain an ionization rate around the field maximum given by

$$\ln W_{ADK}(t) \simeq \ln W_{ADK}(0) + \left(p - \frac{2\kappa^3}{3E_0}\right)\frac{\omega_{UV}^2 t^2}{2}. \tag{23}$$

In the above equation, $p \ll \kappa^3/E_0$, such that it can be neglected and we obtain that around the field maximum

$$W_{ADK}(t) \simeq W_{ADK}(0) \exp\left(-\frac{t^2}{2\sigma_t^2}\right), \tag{24}$$

where the temporal tunneling width is given by

$$\sigma_t = \sqrt{\frac{3}{\kappa^3}}\frac{I_0^{1/4}}{\omega_{UV}}. \tag{25}$$

This width reflects the minimal temporal gating imposed by the UV pulse in the tunneling regime. It increases only slowly with UV intensity and is expected to decrease with higher UV frequency. However, at higher frequencies, multiple peaks can contribute, which may ultimately complicate the structure of the attosecond pulse signal.

## III. WAVELENGTH SCALING OF THE SPECTRAL DENSITY

We repeated the efficiency calculations using the same pulse parameters as in the main text, with the UV pulse at two phase delays, $\phi_d = -7/16$ and $\phi_d = -3/8$. This time, the efficiency was computed over a broader frequency window, extending from the UV cutoff to the end of the plateau, so that no particular frequency region is favored. A fit of the form $\propto \lambda^{-x}$ was then applied. As shown in Fig. 5, the MIR-only case scales approximately as $\lambda^{-4.46}$. Adding the UV pulse increases the yield by about four to five orders of magnitude for UV intensities of $10^{14}$ W/cm$^2$ and $5 \times 10^{14}$ W/cm$^2$, respectively. Moreover, the signals for the two phase delays are now much closer, since each enhances different frequency regions that are both included in the efficiency estimate.

7

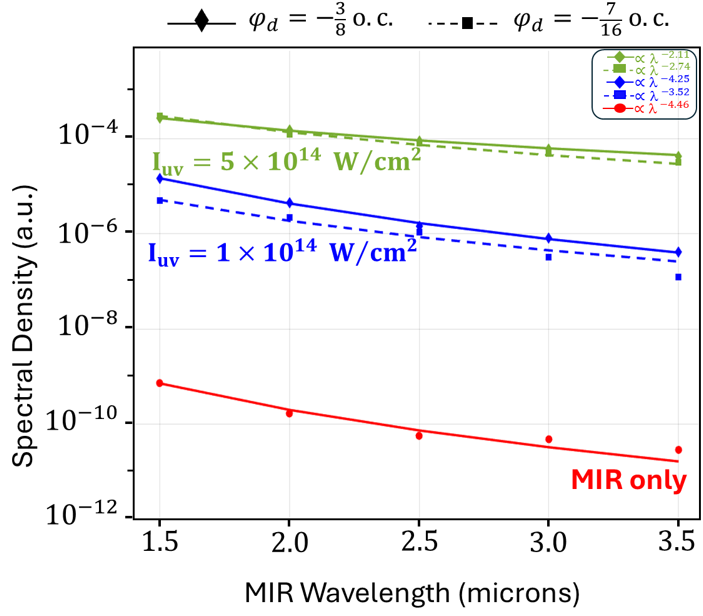

FIG. 5. Efficiency scaling of the HHG plateau for varying wave lengths of the MIR pulse at two different phase delays $\phi_d = -3/8$ o.c. and $\phi_d = -7/16$ o.c. and at two different UV intensities $I_{uv} = 1 \times 10^{14}$ W/cm² and $I_{uv} = 5 \times 10^{14}$ W/cm².

## IV. ISOLATED ATTOSECOND PULSE GENERATION WITH MULTI-CYCLE MIR

We repeated the TDSE calculations with an 8-cycle (FWHM = 30 fs), 3 $\mu$m MIR pulse and analyzed the generated attosecond emission, both for the MIR-only and the MIR+UV schemes. The UV pulse was introduced at $\phi_d = -3/8$ to preferentially select short trajectories. As shown in Fig. 6, the MIR-only case produces three main attosecond bursts, originating primarily from long trajectories at successive MIR peaks. By contrast, when combining the multi-cycle MIR with a UV pulse of $5 \times 10^{14}$ W/cm² and CEP = 90°, we obtain a single attosecond pulse with a duration of 172 as (FWHM). The signal is enhanced by more than five orders of magnitude compared to the MIR-only case, highlighting the robustness of the scheme even for long MIR drivers.

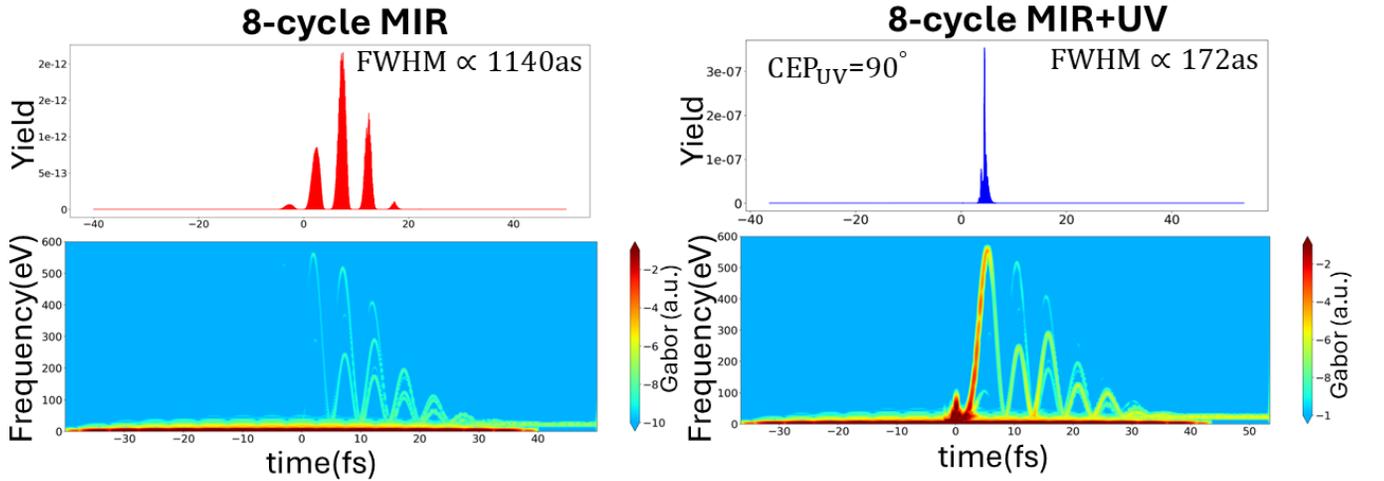

FIG. 6. Attosecond pulse and Gabor transform with an 8-cycle MIR Driver without UV and with UV. The center of the UV pulse is placed at phase delay $\phi_d = -3/8$ o.c.

## V. PHASE MATCHING FOR MACROSCOPIC PULSE PROPAGATION

In this section, we briefly consider the experimental feasibility of producing attosecond pulses in the X-ray regime using the proposed two-color scheme. To realize the two-color driving scheme experimentally, one must consider several key factors associated with the macroscopic propagation of the UV, mid-IR, and soft x-ray pulses. While a full macroscopic calculation is beyond the scope of the present work, analytical estimates of the relevant wave propagation dynamics are suitable to evaluate the experimental feasibility of the approach. First, realizing the observed efficiency enhancement in an extended medium requires phase matching of the generated soft x-ray attosecond pulses with the mid-IR driving field. While a detailed account of phase matching must include an analysis of the experimental geometry, phase matching is possible provided that the ionization fraction $\eta$ is less than the critical ionization level [9], which is $\eta_{crit} \approx 0.0325\%$ for a 3.5-$\mu$m pulse in helium. This requirement thus restricts the UV intensity to approximately $10^{14}$ W/cm$^2$, which corresponds to an enhancement of the harmonic flux by approximately four orders of magnitude in comparison to that of the mid-IR driver only.

Because the ionization must be locked to a particular phase of the mid-IR field as it propagates, we must also evaluate the mismatch between the *group velocity* of the UV pulse and the *phase velocity* of the mid-IR field in the partially ionized medium. These dynamics differ from those of phase matching, since the group velocity must always remain below $c$, while the phase velocity in plasma can be greater than $c$. We find that for $\eta = \eta_{crit}$, the temporal walk-off remains smaller than the UV pulse duration pressure-length products up to several bar-cm, thereby allowing operation of HHG gas cells at multi-atmosphere pressures, as have previously been used for soft x-ray attosecond pulse generation [10, 11].

Finally, we note that the UV RDWs with near-PW/cm$^2$ peak intensities, such as those considered here, have not been experimentally characterized to date. However, RDWs with pulse energies at the 10 $\mu$J level [12] can now be routinely generated from post-compressed Ti:Sapphire lasers, and RDW pulse durations of <1.5 fs have recently been characterized [13], at central wavelengths near the 240 nm chosen for our study. When focused to a relatively modest beam waist of 20 $\mu$m, such pulses could yield intensities well in excess of the $10^{14}$ W/cm$^2$ level required for the highest phase-matched conversion efficiency. Moreover, post-compression [14] and RDW emission [15] have both been shown to scale well to long-wavelength drivers, providing a viable pathway towards the generation of synchronized UV pulses and mid-IR driving fields

---

[1] N. Douguet, A. N. Grum-Grzhimailo, E. V. Gryzlova, E. I. Staroselskaya, J. Venzke, and K. Bartschat, Phys. Rev. A **93**, 033402 (2016), URL https://link.aps.org/doi/10.1103/PhysRevA.93.033402.

[2] N. Douguet, M. Guchkov, K. Bartschat, and S. F. d. Santos, Atoms **12** (2024), ISSN 2218-2004, URL https://www.mdpi.com/2218-2004/12/7/34.

[3] X. M. Tong and C. D. Lin, Journal of Physics B: Atomic, Molecular and Optical Physics **38**, 2593 (2005), URL https://dx.doi.org/10.1088/0953-4075/38/15/001.

[4] A. T. Bondy, S. Saha, J. C. del Valle, A. Harth, N. Douguet, K. R. Hamilton, and K. Bartschat, Phys. Rev. A **109**, 043113 (2024), URL https://link.aps.org/doi/10.1103/PhysRevA.109.043113.

[5] M. Lewenstein, P. Balcou, M. Y. Ivanov, A. L'Huillier, and P. B. Corkum, Phys. Rev. A **49**, 2117 (1994), URL https://link.aps.org/doi/10.1103/PhysRevA.49.2117.

[6] A.-T. Le, H. Wei, C. Jin, and C. D. Lin, Journal of Physics B: Atomic, Molecular and Optical Physics **49**, 053001 (2016), URL https://dx.doi.org/10.1088/0953-4075/49/5/053001.

[7] K. L. Ishikawa, K. Schiessl, E. Persson, and J. Burgdörfer, Phys. Rev. A **79**, 033411 (2009), URL https://link.aps.org/doi/10.1103/PhysRevA.79.033411.

[8] G. L. Yudin and M. Y. Ivanov, Phys. Rev. A **64**, 013409 (2001), URL https://link.aps.org/doi/10.1103/PhysRevA.64.013409.

[9] T. Popmintchev, M.-C. Chen, A. Bahabad, M. Gerrity, P. Sidorenko, O. Cohen, I. P. Christov, M. M. Murnane, and H. C. Kapteyn, Proceedings of the National Academy of Sciences **106**, 10516 (2009), https://www.pnas.org/doi/pdf/10.1073/pnas.0903748106, URL https://www.pnas.org/doi/abs/10.1073/pnas.0903748106.

[10] J. Li, X. Ren, Y. Yin, K. Zhao, A. Chew, Y. Cheng, E. Cunningham, Y. Wang, S. Hu, Y. Wu, et al., Nature Communications **8** (2017), URL https://doi.org/10.1038/s41467-017-00321-0.

[11] S. L. Cousin, N. Di Palo, B. Buades, S. M. Teichmann, M. Reduzzi, M. Devetta, A. Kheifets, G. Sansone, and J. Biegert, Phys. Rev. X **7**, 041030 (2017), URL https://link.aps.org/doi/10.1103/PhysRevX.7.041030.

[12] J. C. Travers, T. F. Grigorova, C. Brahms, and F. Belli, Nat. Photonics **13**, 547–554 (2019), URL https://www.nature.com/articles/s41566-019-0416-4.

[13] A. M. Heinzerling, F. Tani, M. Agarwal, V. S. Yakovlev, F. Krausz, and N. Karpowicz, Nature Photonics **19**, 772 (2025), ISSN 1749-4893, URL https://doi.org/10.1038/s41566-025-01658-5.



\end{document}